    \documentclass[onecolumn]{aa}
    \usepackage{txfonts}
  \usepackage{graphicx}
    \usepackage{ifpdf}
    \usepackage{epsfig}
    \usepackage[latin1]{inputenc}
    \usepackage{amsfonts}
    \usepackage{amssymb}
    \usepackage{latexsym}
    \usepackage{amsbsy}
    \usepackage{amstext}
    \usepackage{amsopn}
    \usepackage{amsgen}
    \usepackage[dvips]{color}
    \usepackage{color}
 \newcommand{\inc}{{\it i}}

 \newcommand{\be}{\begin{equation}}
 \newcommand{\ee}{\end{equation}}
 \newcommand{\ba}{\begin{eqnarray}}
 \newcommand{\ea}{\end{eqnarray}}

 \newcommand{\erbold}{\mbox{{\boldmath $\vec r$}}}

  \newcommand{\eRbold}{\mbox{{\boldmath $\vec{R}$}}}
  \newcommand{\Rbold}{\mbox{{\boldmath $\vec{R}$}}}

\begin{document}
 \title{
 Justification of the two-bulge method in the theory of bodily tides
            }
 %
 %
 \author{Michael Efroimsky
          \inst{1}
          }
 \institute{US Naval Observatory, Washington DC 20392 USA\\
              \email{~michael.efroimsky$\,$@$\,$usno.navy.mil}
                      }

   \date{Received: ~25 April 2012; ~accepted: ... }

 \begin{abstract}
 {}
 {Mathematical modeling of bodily tides can be carried out in various ways. Most straightforward is the method of complex amplitudes,
 which is often used in the planetary science. Another method, employed both in planetary science and in astrophysics, is based on
 decomposition of each harmonic of the tide into two bulges oriented orthogonally to one another. We prove that the two methods are
 equivalent. Specifically, we demonstrate that the two-bulge method is not a separate approximation, but ensues directly from the Fourier
 expansion of a linear tidal theory equipped with an arbitrary rheological model involving a departure from elasticity.}
 {To this end, we use the most general mathematical formalism applicable to linear bodily tides. To express the tidal amendment to
 the potential of the perturbed primary, we act on the tide-raising potential of the perturbing secondary with a convolution
 operator.}
 {This enables us to interconnect a complex Fourier component of the tidally generated potential of the perturbed primary with the
 appropriate complex Fourier component of the tide-raising potential of the secondary. Then we demonstrate how this interrelation
 entails the two-bulge description.}
 {While less economical mathematically, the two-bulge approach has a good illustrative power, and may be employed on a par with a more
 concise method of complex amplitudes. At the same time, there exist situations where the two-bulge method becomes more practical for
 technical calculations.}
 \end{abstract}


 \maketitle

 \section{Introduction and aim}

 On several occasions, it was suggested by different authors to model bodily tides with superposition of two symmetrical bulges. One
 bulge is always aimed at the secondary, and thus implements the instantaneous reaction of the primary's shape and potential to the
 tide-rising gravitational pull exerted by the secondary. This portion of the tide is called ``adiabatic tide" (Zahn 1966a,b) or
 ``elastic tide" (Ferraz Mello 2012; Krasinsky 2006). The second bulge is assumed to align orthogonally to the direction to the
 tide-raising secondary, and thus is set to implement the entire nonelastic portion of the primary's deformation. This, second bulge is
 called ``dissipative tide" (Zahn 1966a,b; Krasinsky 2006) or ``creep tide" (Ferraz Mello 2012).

 In this note, we demonstrate that the two-bulge method is not a separate approximation, but ensues directly from the Fourier expansion
 of a linear tidal theory equipped with an arbitrary rheological model involving a departure from elasticity.

 \section{The static linear theory of bodily tides}

 Let a spherical primary of radius $\,R\,$ be subject to the gravitational pull by a secondary of mass $M_{sec}\,$, residing at
 the position $\,{\erbold}^{\;*} = (r^*,\,\phi^*,\,\lambda^*)\,$, where $\,r^*\geq R\,$. At a surface point $\Rbold = (R,\phi,
 \lambda)$ of the primary body, the tidal potential generated by the secondary can be expanded over the Legendre polynomials $\,P_{\it l}(\cos\gamma)\;$ as
 \ba
 W(\eRbold\,,\,\erbold^{~*})&=&\sum_{{\it{l}}=2}^{\infty}~W_{\it{l}}(\eRbold\,,~\erbold^{~*})~=~-~\frac{G\;M_{sec}}{r^{
 \,*}}~\sum_{{\it{l}}=2}^{\infty}\,\left(\,\frac{R}{r^{\,*}}\,\right)^{\textstyle{^{\it{l}}}}\,P_{\it{l}}(\cos \gamma)~~~.
 \label{1}
 \ea
 Here $\,G\,$ denotes Newton's gravity constant, while $\gamma\,$ is the angular separation between the vectors ${\erbold}^{\;*}$ and
 $\Rbold$ pointing from the primary's centre. The longitudes $\lambda,\,\lambda^*$ are measured from a fixed meridian on the primary
 body, the latitudes $\phi,\,\phi^*$ being reckoned from the equator. The index $\,l\,$ is conventionally named as the {\it{degree}}.
 In (\ref{1}) the $\,l=0\,$ term is missing, because it corresponds to the principal, Newtonian part of the secondary's potential,
 and is not a part of the perturbation. Omission of the $\,l=1\,$ term is a more subtle point related to the fact that we are describing
 the motion of the secondary relative to the primary body, and not relative to an inertial frame (see equations 8 - 11 in
 Efroimsky \& Williams 2009).

 Within the linear theory, the $\,{\emph{l}}^{~th}$ term $\,W_{\it{l}}(\eRbold\,,~\erbold^{~*})\,$ of the secondary's potential
 generates a linear alteration of the primary's shape. This alteration, in its turn, causes a linear amendment $\,U_{\it{l}}(\erbold)\,$
 to the gravitational potential of the primary, where $\,${\it{linear}}$\,$ means: linear in $\,W_{\it{l}}(\eRbold\,,~\erbold^{~*})\,$. The
 theory of potential requires that outside the primary body $\,U_{\it{l}}(\erbold)\,$ should scale with the distance as $\,r^{-(\it{l}+1)}\,$.
 Hence the said change in the primary's potential may be written down as
 \ba
 U_{\it{l}}(\erbold)\,=\,k_l\,\left(\,\frac{R}{r}\,\right)^{{\it l}+1}\;W_{\it{l}}(\eRbold\,,\;\erbold^{\;*})~~~,
 \label{2}
 \ea
 where $R\,$ is the mean equatorial radius of the primary, $\,\eRbold\,=\,(R\,,\,\phi\,,\,\lambda)\,$ is a point on the primary's surface,
 while $\,\erbold\,=\,(r\,,\,\phi\,,\,\lambda)\,$ is an exterior point right above the surface point $\,\eRbold\,$, at a radius $\,r\,\geq\,R\,$.
 The  numerical factors $\,k_l\,$ are the degree-$l$ {\it{Love numbers}} calculated from the rheology of the primary body.

 The overall tidally caused change of the primary's potential thus amounts to
 \ba
 U(\erbold)~=~\sum_{{\it l}=2}^{\infty}~U_{\it{l}}(\erbold)~=~\sum_{{\it l}=2}^{\infty}~k_{\it
 l}\;\left(\,\frac{R}{r}\,\right)^{{\it l}+1}\;W_{\it{l}}(\eRbold\,,\;\erbold^{\;*})~~~.~~~~~~~
 ~~~~~~~~~~~~~~~~
 \label{3}
 \ea

 \section{Dynamical linear theories of bodily tides}

 In realistic situations, the disturbing potential is evolving, so equation (\ref{1}) assumes the form of
 \ba
 W(\eRbold\,,\,\erbold^{~*}(t)\,)&=&\sum_{{\it{l}}=2}^{\infty}~W_{\it{l}}(\eRbold\,,~\erbold^{~*}(t)\,)~=~-~\frac{G\;
 M_{sec}}{r^{\,*}(t)}~\sum_{{\it{l}}=2}^{\infty}\,\left(\,\frac{R}{r^{\,*}(t)}\,\right)^{\textstyle{^{\it{l}}}}\,P_{\it{l}}
 (\cos \gamma(t)\,)~~.~
 \label{4}
 \ea
 Then one should expect the distortion of the primary, as well as the corresponding amendment to its potential at an exterior
 point $\,\erbold\,$, to become a function of time: $\,U(\erbold,\,t)\,$.

 \subsection{Elastic dynamical tides}

 Had the tides contained only instantaneous, elastic components, the expressions for the tidal potential would mimic
 (\ref{2} - \ref{3}). At each instant of time $\,t\,$, the degree-$l$ term of the tide-raising potential $\,W\,$ of the
 orbiting secondary would generate instantaneously an appropriate degree-$l$ term of the tidal potential of the primary:
 \ba
 ^{\textstyle{^{(elastic)}}}U_{\it{l}}(\erbold,\,t)\,=\,k_l\,\left(\,\frac{R}{r}\,\right)^{{\it l}+1}\;W_{\it{l}}(\eRbold
 \,,\;\erbold^{\;*}(t)\,)~~~,
 \label{5}
 \ea
 so the total tidal amendment to the potential of the primary would look:
 \ba
 ^{\textstyle{^{(elastic)}}}U(\erbold,\,t)~=~\sum_{{\it l}=2}^{\infty}~~^{\textstyle{^{(elastic)}}}U_{\it{l}}(\erbold,\,t)
 ~=~\sum_{{\it l}=2}^{\infty}~k_{l}~\left(\,\frac{R}{r}\,\right)^{{\it l}+1}~W_{\it{l}}(\eRbold\,,~\erbold^{~*}(t)\,)~~~.~~~
 \label{6}
 \ea
 Needless to say, in realistic materials the internal friction prevents the tidal deformation from being instantaneous.

 \subsection{Realistic dynamical tides}

 To describe deviation from elasticity, we spell the two basic assumptions
 whereon a linear dynamical theory of bodily tides is based:

 \begin{itemize}
 \item[\bf{[1]}] tidal deformation is linear with respect to the stress generated by the tide-raising potential;\\

 \item[\bf{[2]}] the deformation is not fully elastic: it incorporates both an immediate and delayed portions
 (delayed -- relative to the tide-raising potential).
 \end{itemize}

 Mathematically, assumption {\bf{[1]}} means that an infinitesimal increment $\,\Delta W_{\it{l}}(\eRbold\,,\;\erbold^{\;*}
 (t\,')\,)\,$, whereby the perturbing potential increased at the time $\,t\,'\,$ in the past, results in a proportional
 present-time increment of the tidally distorted shape of the primary and, accordingly, in a proportional increment of the
 tidal amendment to its potential:
 \ba
 \Delta U_{\it l}(\erbold,\,t)~=~\left(\frac{R}{r}\right)^{{\it l}+1}{\it{k}}_{\textstyle{_l}}(t-t\,')~\,\Delta W_{\it{l}}
 (\eRbold\,,\;\erbold^{\;*}(t\,')\,)~~~,
 \label{7}
 \ea
 ${\it{k}}_{\textstyle{_l}}(t-t\,')\,$ being a function describing the delayed reaction of the shape. The mathematical
 form of this function is defined by the rheology of the body and by its self-gravitation.

 Thus the addition $\,U_l\,$ to the primary's potential gets expressed
 through the tide-raising potential $\,W_l\,$ by a linear integral operator:
 \ba
 U_{\it l}(\erbold,\,t)\,=\,\left(\frac{R}{r}\right)^{{\it l}+1}\int_{t\,'=-\infty}^{t\,'=t} k_{\it l}(t-t\,')\stackrel{\bf
 \centerdot}{W}_{\it{l}}(\eRbold\,,\,\erbold^{\;*}(t\,')\,)\,dt\,'~~,
 \label{8a}
 \ea
 overdot denoting a time derivative. Integration of (\ref{8a}) by parts renders:
 \ba
 U_{\it l}(\erbold,\,t)\,=\,\left(\frac{R}{r}
 \right)^{{\it l}+1}\left[k_l(0)W_l(t)\,-\,k_l(\infty)W_l(-\infty)\right]\,+\,\left(\frac{R}{r}
 \right)^{{\it l}+1}\int_{-\infty}^{t}
 {\bf\dot{\it{k}}}_{\textstyle{_l}}(t-t\,')\,~W_{\it{l}}
 (\eRbold\,,\,\erbold^{\;*}(t\,')\,)\,dt\,'~\,,~\quad
 \label{8b}
 \ea
 where the relaxed term $\,-\,k_l(\infty)W(-\infty)\,$ should be neglected. Indeed, the current events cannot be influenced
 by the perturbation $\,W_l(-\infty)\,$ in the infinite past, wherefore $\,k_l(\infty)=0\,$. Of the remaining two terms,
 the unrelaxed term $\,k_l(0)W_l(t)\,$ reflects the elastic part of the deformation, while the integral expresses the
 delayed components -- viscous and anelastic.

 Assumption {\bf{[2]}} means that both the unrelaxed term $\,k_l(0)W_l(t)\,$ and the delayed term given by the integral
 should be kept. As explained in Efroimsky (2012a,b), the relaxed term may be easily incorporated into the integral, where
 it should show up multiplied with a Heaviside step function $\,\Theta(t-t\,')\,$. Then our expression for the tidal potential
 will acquire the simple form of
 \ba
 U_{\it l}(\erbold,\,t)\;=\;\left(\frac{R}{r}
 \right)^{{\it l}+1}\int_{-\infty}^{t} {\stackrel{\bf\centerdot}{\it{k}}}_{\textstyle{_l}}(t-t\,')~W_{\it{l}}
 (\eRbold\,,\;\erbold^{\;*}(t\,')\,)\,dt\,'~,
 \label{9}
 \ea
 ${\bf{\it{k}}}_{\textstyle{_l}}(t-t\,')$ now incorporating both the delayed-reaction terms and the
 elastic term $\,{k}_{l}(0)\,\Theta(t-t\,')\,$. The elastic part will enter the kernel $\,{\stackrel{\bf\centerdot}{\it{
 k}}}_{\textstyle{_l}}(t-t\,')\,$ as $\,{k}_{l}(0)\,\delta(t-t\,')\,$, with $\,\delta(t-t\,')\,$ being the Dirac delta
 function. Integration of this term will furnish $\,k_l(0)W_l(t)\,$, as in (\ref{8b}).

 \section{Fourier components of tidal stresses and strains}

 \subsection{Tidal modes and tidal frequencies}

 The sidereal angle and the spin rate of a tidally-perturbed primary are normally denoted with $\,\theta\,$ and $\,\stackrel{\bf
 \centerdot}{\theta\,}\,$, while the node, pericentre, and mean anomaly of a tide-raising secondary, as seen from the primary, are
 denoted with $\,\Omega\,$, $\omega\,$, and ${\cal M}$.

 In the Darwin-Kaula theory, the tide-raising potential $W$, the primary's deformation, and the tidal amendment $U$ to the primary's
 potential are expanded over the modes
 \ba
 \omega_{lmpq}\;\equiv\;(l-2p)\;\dot{\omega}\,+\,(l-2p+q)\;\dot{\cal{M}}\,+\,m\;(\dot{\Omega}\,-\,\dot{\theta})\,\approx\,
 (l-2p+q)\;n\,-\,m\;\dot{\theta}
 ~~~,~~~
 \label{10}
 \ea
 $l,\,m,\,p,\,q\,$ being integers, and $\,n\,$ being the mean motion. Dependent upon the values of the mean motion, spin rate, and the indices, the tidal modes
 $\,\omega_{{\it l}mpq}\,$ may be positive or negative or zero.

 The actual forcing frequencies of the resulting stresses and strains in the primary's material are the absolute values of the
 tidal modes:
 \ba
 \label{11}
 \chi_{lmpq}\equiv\,|\,\omega_{lmpq}\,|~~~,
 \ea
 so these frequencies are always positive.

 \subsection{Fourier expansions}

 In practical calculations, it is extremely convenient to employ complex stresses and strains, under the convention that the actual,
 physical quantities are the real parts of their complex counterparts. This way, the Fourier series for the stress and strain look:
 \ba
 \sigma_{\gamma\nu}(t)\,
 \,=\,\sum_{s=0}^{\infty}\,\sigma_{\gamma\nu}(\chi_{\textstyle{_s}})\,\cos\left[\,\chi_{\textstyle{_s}}
 t+\varphi_{\sigma}(\chi_{\textstyle{_s}})\,\right]\;
 =\;\sum_{s=0}^{\infty}\,
 {\cal{R}}{\it{e}}\left[\,{\bar{\sigma}}_{\gamma\nu}(\chi_{\textstyle{_s}})\,\;\exp\left({\textstyle{{\,\inc
 \chi_{\textstyle{_s}} t}}}\right)\,\;\right]~~~,~~~~~~~~~~
 \label{12}
 \ea
 \ba
 u_{\gamma\nu}(t)
 \,=\,\sum_{s=0}^{\infty}\,u_{\gamma\nu}(\chi_{\textstyle{_s}})\,\cos\left[\,\chi_{\textstyle{_s}}t+
 \varphi_{u}(\chi_{\textstyle{_s}})\,\right]
 \;
 =\;\sum_{s=0}^{\infty}\,
 {\cal{R}}{\it{e}}\left[\,{\bar{u}}_{\gamma\nu}(\chi_{\textstyle{_s}})\;\,\exp\left({\textstyle{{\,
 \inc\chi_{\textstyle{_s}} t}}}\right)\,\;\right]~~~,~~~~~~~~~~
 \label{13}
 \ea
 $\gamma\nu$ being tensor indices, and $s$ being a concise notation for $lmpq$. The complex amplitudes are
 \ba
 {\bar{{\sigma}}_{\gamma\nu}}(\chi)={{{\sigma}}_{\gamma\nu}}(\chi)\,\;\exp\left[{\inc\varphi_\sigma(\chi)}\right]~~~~~,
 ~~~~~~{\bar{{u}}_{\gamma\nu}}(\chi)={{{u}}_{\gamma\nu}}(\chi)\,\;\exp\left[{\inc\varphi_u(\chi)}\right]~~~,
 \label{compamp}
 \label{14}
 \ea
 where the initial phases $\,\varphi_{\sigma}(\chi)\,$ and $\,\varphi_{u}(\chi)\,$ are chosen so that the real amplitudes
 $\,\sigma_{\gamma\nu}(\chi_{\textstyle{_s}})\,$ and $\,u_{\gamma\nu}(\chi_{\textstyle{_s}})\,$ are non-negative.

 For a continuous spectrum, the sums get replaced with integrals over frequency:
 \ba
 \sigma_{\gamma\nu}(t)~=~{\cal{R}}{\it{e}}~\int_{0}^{\infty}\,\bar{\sigma}_{\gamma\nu}(\chi)~e^{\textstyle{^{\,\inc\chi t}}}~
 d\chi\quad\quad\mbox{and}~\quad~\quad
 u_{\gamma\nu}(t)~=~{\cal{R}}{\it{e}}~\int_{0}^{\infty}\,\bar{u}_{\gamma\nu}(\chi)~e^{\textstyle{^{\,\inc\chi t}}}~d\chi~~~.
 \label{15}
 \ea
 Similarly, the tide-raising potential $W_l$ and the potential $U_l$ of the primary get expanded into a sum or integral
 over the tidal modes:
 \ba
 W_{l}(t)~=~{\cal{R}}{\it{e}}~\int_{-\infty}^{\infty}\,\bar{W}_{l}(\omega)~e^{\textstyle{^{\,\inc\omega t}}}~d\omega\quad\quad
 \mbox{and}~\quad~\quad
 U_{l}(t)~=~{\cal{R}}{\it{e}}~\int_{-\infty}^{\infty}\,\bar{U}_{l}(\omega)~e^{\textstyle{^{\,\inc\omega t}}}~d\omega~~~,
 \label{16}
 \ea
 where the complex amplitudes are expressed via the real amplitudes and the initial phases by
 \ba
 {\bar{{W}}_{l}}(\omega)={{{W}}_{l}}(\omega)\,\;\exp\left[{\inc\varphi_{\textstyle{_{W_l}}}(\omega)}\right]~~~~~,
 ~~~~~~{\bar{{U}}_{l}}(\omega)={{{U}}_{l}}(\omega)\,\;\exp\left[{\inc\varphi_{\textstyle{_{U_l}}}(\omega)}\right]~~~.
 \label{compamp}
 \label{17}
 \ea
 The phases $\,\varphi_{\textstyle{_{W_l}}}(\omega)\,$ and $\,\varphi_{\textstyle{_{U_l}}}(\omega)\,$ can always be set in such a
 way that the real amplitudes $\,W_{l}(\chi)\,$ and $\,U_{l}(\chi)\,$ are non-negative.

 Both in (\ref{15}) and (\ref{16}),  the actual, physical spectral components are the real parts of the complex ones. However, there
 also is an important difference between (\ref{15}) and (\ref{16}). While the stresses and strains are habitually expanded in
 (\ref{15}) over positive frequencies $\,\chi\,$ only, the potentials in (\ref{16}) are expanded over the tidal modes $\,\omega\,$,
 which can be positive or negative or zero, as demonstrated in the Darwin-Kaula theory of tides.

 It is of course a common fact that a real function can be decomposed into a Fourier series or integral over only positive
 frequencies. This way, expansions of the potentials over $\,\chi\equiv | \omega | \,$ may appear to be sufficient, because a
 contribution from some negative tidal mode $\,\omega<0\,$ can be shown to coincide with the contribution from the appropriate
 positive mode $\,|\omega|>0\,$. Therefore, (\ref{16}) may be rewritten simply as
 \ba
 W_{l}(t)~=~{\cal{R}}{\it{e}}~\int_{0}^{\infty}\,\bar{W}_{l}(\chi)~e^{\textstyle{^{\,\inc\chi t}}}~d\chi\quad\quad\mbox{and}
 ~\quad~\quad
 U_{l}(t)~=~{\cal{R}}{\it{e}}~\int_{0}^{\infty}\,\bar{U}_{l}(\chi)~e^{\textstyle{^{\,\inc\chi t}}}~d\chi~~~,
 \label{18}
 \ea
 where $\,\bar{W}_{l}(\chi)\,=\,2\,\bar{W}_{l}(\omega)\,$ and $\,\bar{U}_{l}(\chi)\,=\,2\,\bar{U}_{l}(\omega)\,$.

 Surprisingly, the theory of tides is a rare exception from the rule, in that this theory does distinguish between the contribution
 from a negative tidal mode and that from a positive mode of the same absolute value. Fortunately, the difference shows up only at
 the stage when one calculates tidal forces or torques (Efroimsky 2012a,b). As in the current paper we discuss potentials only, we
 shall ignore this subtlety and shall employ (\ref{18}) instead of (\ref{16}).

 \subsection{Dynamical analogues to the Love number}

 Insertion of (\ref{18}) into (\ref{9}) entails
 \ba
 \bar{U}_{\textstyle{_{l}}}(\chi)\;=\;\left(\frac{R}{r}\right)^{l+1}\bar{k}_{\textstyle{_{l}}}(\chi)\;\,\bar{W}_{\textstyle{_{l}}}(\chi)
 \label{19a}
 \ea
 or, in a more detailed form:
 \ba
 |\,\bar{U}_{\textstyle{_{l}}}(\chi)\,|~e^{\,\textstyle{^{i\,\chi\,t\,+\,i\,\varphi_{\textstyle{_{_{U_l}}}}(\chi)}}}~=~
 \left(\frac{R}{r}\right)^{l+1}~
 |\,\bar{k}_{\textstyle{_{l}}}(\chi)\,|\;~e^{\,\textstyle{^{i\,\chi\,t\,-\,i\,\epsilon_l(\chi)}}}\;
 |\,\bar{W}_{\textstyle{_{l}}}(\chi)\,|\;~e^{\,\textstyle{^{i\,\chi\,t\,+\,i\,\varphi_{\textstyle{_{_{W_l}}}}(\chi)}}}\;\;\;,
 \label{19b}
 \label{VR}
 \ea
 where the complex function
 \ba
 \bar{k}_{\textstyle{_{l}}}(\chi)\,=\,|\bar{k}_{\textstyle{_{l}}}(\chi)|~e^{\textstyle{^{\,-\,i\,\epsilon_l(\chi)\,}}}~~~,
 \label{20}
 \ea
 is a Fourier component of the kernel $\,{\bf\dot{\it{k}}}_{\textstyle{_l}}(t-t\,')\,$ of the integral operator (\ref{9}). The
 kernels $\,{\bf\dot{\it{k}}}_{\textstyle{_l}}(t-t\,')\,$ are named {\it{Love functions}}, a term suggested by Churkin (1998). The
 functions $\,\bar{k}_{\textstyle{_{l}}}(\chi)\,$ are called {\it{complex Love numbers}}, in understanding however that these
 ``numbers" change with frequency. An approach similar to (\ref{20}) was taken by Mathis and Le Poncin Lafitte (2009). These authors
 introduced a complex impedance as the ratio between the complex Love number and the static Love number. Their equation (107) is
 equivalent to our (\ref{20}), up to a convention on the sign of the argument.

 We see from (\ref{19b}) that at each frequency $\,\chi\,$, the negative argument $\,\epsilon_l(\chi)~$ is a measure of lagging of the
 spectral component $\,U_l(\chi)\,$, relative to appropriate spectral component $\,W_l(\chi)~$:
 \ba
 \varphi_{\textstyle{_{U_l}}}(\chi)~=~\varphi_{\textstyle{_{W_l}}}(\chi)~-~\epsilon_l(\chi)~~.
 \label{21}
 \ea

 \section{Decomposition of a tidal mode into an in-phase part and an in-quadrature (lagging by $\bf{90^o}$) part}

 In the expression (\ref{19b}), we may set the initial phase of $\,\bar{W}_l(\chi)\,$ to be zero, and reckon the phase of
 $\,\bar{U}_l(\chi)\,$ from $\,\bar{W}_l(\chi)\,$. This will enable us to single out, in $\,\bar{U}_l(\chi)\,$, a part which is in
 phase with $\,\bar{W}_l(\chi)\,$, and also to see what part of $\,\bar{U}_l(\chi)\,$ is out of phase with $\,\bar{W}_l(\chi)\,$.

 According to (\ref{21}), nullification of the phase of $\,\bar{W}_l(\chi)\,$ makes the phase of $\,\bar{U}_l(\chi)\,$ equal to the
 negative argument of $\,\bar{k}_l(\chi)\,$. So (\ref{19b}) will assume the form of
 \ba
 U_{l{\textstyle{_{\,0}}}}(\chi)~e^{\,\textstyle{^{i\,\chi\,t\,-\,i\,\epsilon_l(\chi)}}}~=~
 \left(\frac{R}{r}\right)^{l+1}~
 k_{l{\textstyle{_{\,0}}}}(\chi)\;~e^{\,\textstyle{^{\,-\,i\,\epsilon_l(\chi)}}}\;
 W_{l{\textstyle{_{\,0}}}}(\chi)\;~e^{\,\textstyle{^{i\,\chi\,t}}}\;\;\;,
 \label{22}
 \ea
 where we introduced simplified notations $U_{l{\textstyle{_{\,0}}}}(\chi)\,$,
 $\,W_{l{\textstyle{_{\,0}}}}(\chi)\,$ and $\,k_{l{\textstyle{_{\,0}}}}\,$ for real amplitudes:
 \ba
 U_{l{\textstyle{_{\,0}}}}(\chi)~\equiv~|\,\bar{U}_l(\chi)\,|
 \,\quad,\quad\quad\,
 W_{l{\textstyle{_{\,0}}}}(\chi)~\equiv~|\,\bar{W}_l(\chi)\,|
 \,\quad,\quad\quad\,
 k_{l{\textstyle{_{\,0}}}}(\chi)~\equiv~|\,\bar{k}_l(\chi)\,|~~~.
 \label{23}
 \ea
 By means of the Euler formula, (\ref{22}) can be trivially expanded as
 \ba
 \nonumber
 U_{l{\textstyle{_{\,0}}}}(\chi)~\left[\, \cos\left(\chi t - \epsilon_l(\chi)\,\right) + i \sin\left(\chi t - \epsilon_l(\chi)\,
 \right) \,\right]~\quad~\quad~\quad~\quad~\quad~\quad~\quad~\quad~\quad~\quad~~~~~\\
 \nonumber\\ =~\left(\frac{R}{r}\right)^{2l+1}  k_{l{\textstyle{_{\,0}}}}(\chi)~
 \left[ \,\cos\epsilon_l(\chi) - i \sin\epsilon_l(\chi) \,\right] ~W_{l{\textstyle{_{\,0}}}}(\chi)~\left[\, \cos(\chi t) + i \sin(
 \chi t)\,\right]~~~.
 \label{24}
 \ea
 The actual, physical tide-raising potential $W_l(\chi)$ is the real part of the complex $\bar{W}_l(\chi)$. So it is rendered by
 $\,{W_l}_0(\chi) \cos(\chi t)\,$, where we set the initial phase nil, as agreed above. Similarly, the actual, physical tidal
 potential $U_l(\chi)$ is the real part of the complex $U_l(\chi)$, and it reads as:
 \ba
 \nonumber
 {U_l}_0(\chi)  ~\cos\epsilon_l(\chi)~ \cos(\chi t) -  {U_l}_0(\chi)  ~\sin\epsilon_l(\chi)~ \sin(\chi t)
 ~\quad~\quad~\quad~\quad~\quad~\quad~\quad~\quad~\quad~\quad~~~~~\\
 \nonumber\\
 =\,
 \left(\frac{R}{r}\right)^{2l+1}  {k_l }_0(\chi)  ~\cos\epsilon_l(\chi)  ~{W_l}_0(\chi) ~\cos(\chi t) \,+\,
 \left(\frac{R}{r}\right)^{2l+1}  {k_l }_0(\chi)  ~\sin\epsilon_l(\chi)  ~{W_l}_0(\chi) ~\sin(\chi t)~\,.~~
 \label{25}
 \ea
 In this expression for the actual real $\,U_l\,$, we see not one but two terms. One is the elastic part of the tide, a part that is
 in phase with the real $\,W_l\,$. This is the term proportional to $\,\cos(\chi t)~$:
 \ba
 ^{\textstyle{^{(in~phase)}}}{U_l}~=~{U_l}_0(\chi)~\cos\epsilon_l(\chi)~\cos(\chi t) = \left(\frac{R}{r}\right)^{2l+1}{k_l }_0(\chi)~\cos\epsilon_l(\chi)
 ~{W_l}_0(\chi)~\cos(\chi t)~~~,~~~~
 \label{26}
 \ea
 We see that the ``response factor" is equal to  ${k_l }_0(\chi)  \cos\epsilon_l(\chi)~$,
 where ${k_l }_0(\chi)$  is the real amplitude of the complex Love number (call this amplitude {\it{dynamical Love number}}).

 The second component is the in-quadrature term, which is proportional to $\sin(\chi t)~$:
 \ba
 ^{\textstyle{^{(in~quadrature)}}}{U_l}~=~-~{U_l}_0(\chi)~\sin\epsilon_l(\chi)~\sin(\chi t)=\left(\frac{R}{r}\right)^{2l+1}{k_l }_0(\chi)~\sin\epsilon_l(\chi)
 ~{W_l}_0(\chi)~\sin(\chi t)~\,.~~~
 \label{27}
 \ea
 Naturally, the expression of this term via $W_l$ contains ${k_l }_0(\chi)  \sin\epsilon_l(\chi)~$.

 Therefore, as soon as we express $U_l$ via $W_l$ by an integral operator permitting delayed action (Eqn. 9), we automatically
 arrive at the two components of the tidal $U_l$, at each frequency involved, -- the adiabatic component and the dissipative
 component (using the terms of Zahn 1966) or the elastic one and the creep one (as Ferraz-Mello 2012 named them).\footnote{
 Also see the work by Krasinsky (2006), who employed the terms {\it{elastic}} and {\it{dissipative}}.}

 Similar decomposition will take place for the harmonic modes of the primary's surface
 elevation, except that the Love number $\,h_l\,$ will be involved instead of $\,k_l\,$. For
 the principal, $\,\{lmpq\}=\{2200\}\,$, tidal mode, this situation is illustrated, in a
 very exaggerated manner, by Figure 1.
 \begin{figure}[h]
 \hspace{0.5cm}
  {\includegraphics[width=15.1cm]{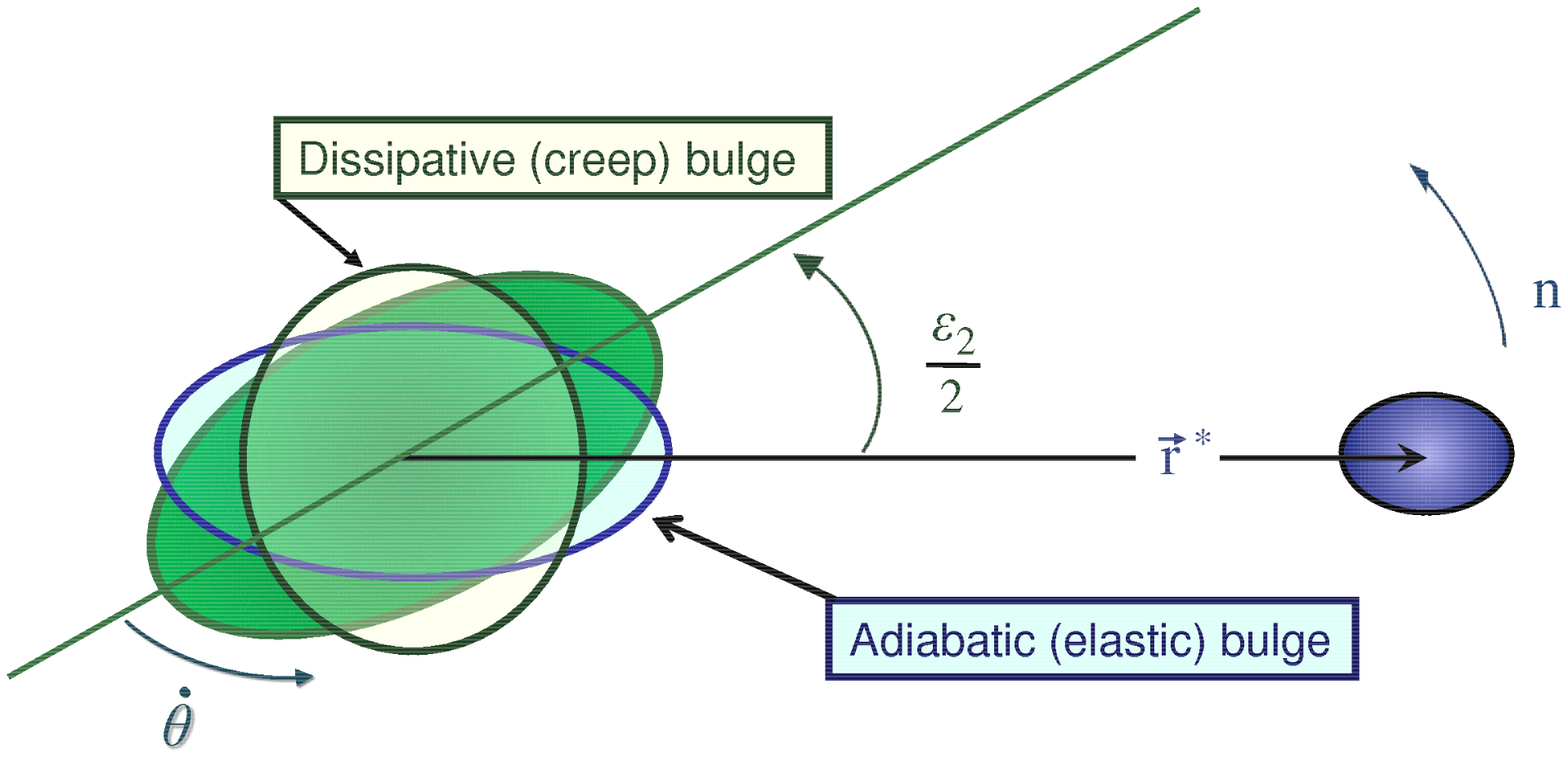}}
 {\caption{\small{Decomposition of the semidiurnal tide into the elastic and dissipative
 components.~\quad~\quad~\quad~\quad~\quad~\quad~\quad~\quad~\quad~\quad~\quad~\quad~\quad~\quad~\quad~\quad~\quad~\quad
 \hspace{0.2cm}
 \vspace{0.3cm}
 \hspace{0.2cm}
 The primary is spinning at the rate $\,\dot{\theta}\,$, with $\,\theta\,$ being its sidereal
 angle. The secondary is orbiting the primary with the mean motion $\,n\,$. The ellipse shaded
 green depicts the principal, semidiurnal mode of the tides exerted on the primary by the
 secondary. This is the mode numbered with $\,\{lmpq\}=\{2200\}\,$. The semimajor axis of the green
 ellipse deviates from the direction towards the perturber by a geometric lag angle. For the
 semidiurnal tide, this angle is equal to $\,\frac{\textstyle 1}{\textstyle 2}\,\epsilon_2(
 \omega_{2200})\,$, where $\,\epsilon_2\,$ is the phase lag as a function of the principal
 tidal mode $\,\omega_{2200}\,$.~\quad~\quad~\quad~\quad~\quad~\quad~\quad~\quad~\quad~\quad~\quad~\quad~\quad~\quad~\quad~\quad~\quad~\quad
  \hspace{6.2cm}
 \vspace{0.3cm}
 \hspace{6.2cm}
 It is shown here, in an exaggerated manner, how the principal tidal mode gets decomposed into
 two components. One is the elastic bulge depicted by the ellipse aligned with the perturber.
 Another is the dissipative bulge depicted with the ellipse aligned orthogonally.
 }}}
 \label{Figure}
 \vspace{0.1cm}
 \end{figure}

 It should finally be mentioned that in the case of stars and giant planets it is, technically, difficult to solve the
 emerging hydrodynamical problem analytically. Therefore, in practical calculations of fluid equilibrium tides in such
 bodies, the two-bulge method is the only known elegant option of getting an acceptable analytical solution.
 One first solves the hydrodynamical equations governing the adiabatic component (\ref{24}). The adiabatic adjustment of
 the structure is deduced from the hydrostatic balance. Then, since the velocity field is preserved along isobars, one
 obtains the radial component of the adiabatic velocity field induced by the tide. The subsequent calculation of the
 horizontal component is based on the velocity field being divergence-free. The knowledge of the adiabatic velocity field
 makes it possible to determine the viscous force. This force drives the dissipative component of the velocity field. The
 process leads to mass redistribution inside the body, and thus to perturbation of the density and gravitational potential
 (\ref{25}). The described, two-bulge approach is available because in the fluid equilibrium tide the dissipative component
 is much weaker than the adiabatic component. The method is implemented, e.g., in the work by Remus et al. (2012a), which
 furthers the original two-bulge approach offered by Zahn (1966).


 \section{Conclusions}

 In this short note, we pointed out a simple rule linking the two methods (or, possibly better to say, two languages), in which
 bodily tides have been described by different authors. The language of complex amplitudes is more economical and is conventional
 to those who studied the theory of vibrations, in physics or engineering. The language of two bulges turns out to be equivalent
 to that of complex amplitudes. Although less economical mathematically, the two-bulges language has some illustrative power, and
 may be employed (like, e.g., in Remus et al. 2012a,b) on a par with the more concise method of complex amplitudes.

 Importantly, the existence of two mutually orthogonal bulges at each tidal frequency is not a separate approximation (as was
 presumed by some of the devotees of the two-bulge method), but is a consequence of the linearity assumption implemented by
 the integral operator (\ref{9}). As soon as we say that the tidal deformation is linear but not fully elastic -- this deformation
 can be decomposed, at each tidal frequency, into an in-phase and an in-quadrature part.

 \section{Acknowledgements}

 It is my pleasure to thank Françoise Remus for her useful advises and for her help in preparing the figure. I also acknowledge
 with gratitude the stimulating conversations on the topic of this work, which I had on various occasions with Sylvio Ferraz Mello,
 Val\'ery Lainey, Valeri Makarov, Stephane Mathis, and Jean-Paul Zahn.
 ~\\

 \section*{}
 {}


\begin{thebibliography}{}


 \bibitem{churkin} Churkin, V. A. 1998. ``The Love numbers for the models of inelastic Earth."
            ~~Preprint No 121. Institute of Applied Astronomy. ~St.Petersburg, Russia. ~/in Russian/

 \bibitem{E&W2009} Efroimsky, M., and Williams, J. G. 2009. ``Tidal torques. A critical review of some techniques."
            {\it{Celestial mechanics and Dynamical Astronomy,}} Vol. {\bf{104}}, pp. 257 - 289.
            ~~~ doi: 10.1007/s10569-009-9204-7 ~~~
            arXiv:0803.3299

 \bibitem{E2012a} Efroimsky, M. 2012a. ``Bodily tides near spin-orbit resonances."
            {\it{Celestial mechanics and Dynamical Astronomy,}}
            Vol. {\bf{112}}, pp. 283 - 330. ~~
            doi: 10.1007/s10569-011-9397-4~.
            ~~Extended version available at ~~ arXiv:1105.6086

 \bibitem{E2012b} Efroimsky, M. 2012b. ``Tidal dissipation compared to seismic dissipation: in small bodies, earths,
            and superearths." {\it{the Astrophysical Journal}}, Vol. {\bf{746}}, No 2, article id 150. ~~
            doi: 10.1088/0004-637X/746/2/150 ~~~  arXiv:1105.3936

 \bibitem{Ferraz} Ferraz-Mello, S. 2012. ``Tidal synchronisation of close-in satellites and exoplanets. A rheophysical approach."
            Submitted to: {\emph{Celestial Mechanics and Dynamical Astronomy}}
            . ~~ arXiv:1204.3957


 \bibitem{Mathis} Mathis, S., and Le Poncin-Lafitte, C. 2009. ``Tidal dynamics of extended bodies in planetary systems and multiple
            stars." {\it{Astronomy \& Astrophysics}}, Vol. {\bf{497}}, pp. 889 -- 910

 \bibitem{Krasinsky} Krasinsky, G. A. 2006. ``Numerical theory of rotation of the deformable Earth with the two-layer fluid core. Part 1:
            Mathematical model." {\emph{Celestial Mechanics and Dynamical Astronomy}}, Vol. {\bf{96}}, pp. 169 - 217. ~~~
            $\left[\right.$In this paper, see formulae (96 - 97) and the subsequent sections 4.2 -- 4.3.$\left.\right]$

 \bibitem{Remus2012a} Remus, F.; Mathis, S.; and Zahn, J.-P. 2012a. ``The Equilibrium Tide in Stars and Giant Planets. I - The Coplanar Case."
           {\it{Astronomy \& Astrophysics}}. In press.

 \bibitem{Remus2012b} Remus, F.; Mathis, S.; Zahn, J.-P.; and Lainey, V. 2012b. ``Anelastic tidal dissipation in multi-layer planets."
            {\it{Astronomy \& Astrophysics}}. In press.

 \bibitem{Zahn1966a} Zahn, J.-P. 1966a. ``Les marées dans une étoile double serrée." {\it{Annales d'Astrophysique,}} Vol. 29, pp. 313 - 330

 \bibitem{Zahn1966b} Zahn, J.-P. 1966b. ``Les marées dans une étoile double serrée (suite)." {\it{Annales d'Astrophysique,}} Vol. 29, pp. 489 - 506

 \end{thebibliography}
 \end{document}